\documentclass[conference]{IEEEtran}
\IEEEoverridecommandlockouts
\usepackage{cite}
\usepackage{amsmath,amssymb,amsfonts}
\usepackage{mathrsfs}
\usepackage{algorithm}
\usepackage{algpseudocode}
\usepackage{amsmath}
\usepackage{graphicx}
\graphicspath{{./fig/}}
\usepackage{textcomp}
\usepackage{xcolor}
\usepackage{enumerate}
\usepackage{bm}
\usepackage{multirow}
\usepackage{makecell}
\usepackage{booktabs}
\usepackage{epstopdf}
\usepackage{caption}
\usepackage{marvosym}
\usepackage{subfig}
\usepackage{float}
\usepackage{soul}
\def\BibTeX{{\rm B\kern-.05em{\sc i\kern-.025em b}\kern-.08em
    T\kern-.1667em\lower.7ex\hbox{E}\kern-.125emX}}
\begin{document}

\title{AdvParams: An Active DNN Intellectual Property Protection Technique via Adversarial Perturbation Based Parameter Encryption}
\author{\IEEEauthorblockN{Mingfu Xue\textsuperscript{1},
        Zhiyu Wu\textsuperscript{2},
        Jian Wang\textsuperscript{1},
        Yushu Zhang\textsuperscript{1},
        and Weiqiang Liu\textsuperscript{3}
      }
\IEEEauthorblockA{\textsuperscript{1}College of Computer Science and Technology, Nanjing University of Aeronautics and Astronautics, Nanjing, China}
\IEEEauthorblockA{\textsuperscript{2}College of Science, Nanjing University of Aeronautics and Astronautics, Nanjing, China}
\IEEEauthorblockA{\textsuperscript{3}College of Electronic and Information Engineering, Nanjing University of Aeronautics and Astronautics, Nanjing, China}
}

\maketitle

\begin{abstract}
The construction of Deep Neural Networks (DNNs) models require high cost, which makes a well-trained DNN model can be regarded as an intellectual property (IP) of the model owner.
To date, many DNN IP protection methods have been proposed, but most of them are watermarking based verification methods where model owners can only verify their ownership passively after the copyright of DNN models has been infringed.
In this paper, we propose an effective framework to actively protect the DNN IP from infringement.
Specifically, we encrypt the DNN model's parameters by perturbing them with well-crafted adversarial perturbations.
With the encrypted parameters, the accuracy of the DNN model drops significantly, which can prevent malicious infringers from using the model.
After the encryption, the positions of encrypted parameters and the values of the added adversarial perturbations form a secret key.
Authorized user can use the secret key to decrypt the model.
Compared with the watermarking methods which only passively verify the ownership after the infringement occurs, the proposed method can prevent infringement in advance.
Moreover, compared with most of the existing active DNN IP protection methods, the proposed method does not require additional training process of the model, which introduces low computational overhead.
Experimental results show that, after the encryption, the test accuracy of the model drops by 80.65\%, 81.16\%, and 87.91\% on Fashion-MNIST, CIFAR-10, and GTSRB, respectively.
Moreover, the proposed method only needs to encrypt an extremely low number of parameters, and the proportion of the encrypted parameters of all the model's parameters is as low as 0.000205\%.
The experimental results also indicate that, the proposed method is robust against model fine-tuning attack and model pruning attack.
Moreover, for the adaptive attack where attackers know the detailed steps of the proposed method, the proposed method is also demonstrated to be robust.
\end{abstract}


\section{Introduction}
Deep Neural Networks (DNNs) have been more and more commercialized because of their excellent performance.
With the emerging of machine learning as a service (MLaaS) \cite{RibeiroGC15}, many commercial companies upload their trained high-performance models to cloud and provide services to the public.
The users can only obtain the model's predictions, but have no access to its internal parameters.
However, training a DNN model with high accuracy is a considerably costly and time-consuming task.
Some malicious infringers may illegally duplicate or abuse the well-trained model, and obtain benefits from it, which greatly infringes the intellectual property (IP) of the model owner.

Most of the existing DNN IP protection methods are watermarking based verification methods.
In these watermarking based verification methods, the DNN model is embedded with watermark, which can be extracted to verify the ownership of the model.
However, these DNN watermarking methods are passive verification methods, which can not prevent the unauthorized usage of the DNN models and can only be applied after the models have already been pirated.

To date, a few active authorization control methods \cite{pyone2020training}, \cite{ChenW18}, \cite{FanNC19}, \cite{lin2020chaotic} have been proposed.
In these methods, only authorized users can normally use the DNN models, while the functionality of the models will be deteriorated seriously for unauthorized users.
Pyone \textit{et al}\cite{pyone2020training} use the secret key to preprocess the training images, then trains a model with these preprocessed training images.
The model only works when the input sample is preprocessed by the secret key.
Chen and Wu\cite{ChenW18} utilizes a data transformation module to realize the function of access control.
However, the works \cite{pyone2020training} and \cite{ChenW18} both require to train the model from scratch, which is a time-consuming process.
Moreover, in the works \cite{pyone2020training} and \cite{ChenW18}, each input sample needs to be transformed before being input into the model, which introduces high computational overhead especially when there are a substantial amount of input samples.
Fan \textit{et al}\cite{FanNC19} embeds the predefined passports to the DNN model so that the model's performance will be deteriorated significantly if the passports are not presented.
The implementation of the work \cite{FanNC19} is time-consuming as it requires to embed the passports into multiple layers of DNN model by retraining the model.
Besides, attackers can compute the hidden parameters by conducting the reverse-engineering attack.

An intuition to realize the active authorization control is to make the model dysfunctional by encrypting the model's parameters.
However, a trained model usually contains a tremendously large number of parameters, which makes the computational overhead of the parameter encryption unacceptable and also substantially enlarges the decryption time.
Thus, it's necessary to search for a few parameters that lead to the most significant deterioration in the performance of the model.
In this paper, we propose an adversarial perturbation based method to actively protect the DNN IP with negligible computational overhead.
We first utilize the gradient of the model's loss function to select the parameters that have the greatest impact on model's performance, then modify the selected parameters with adversarial perturbations.
With an extremely low number of encrypted parameters, the model will be dysfunctional significantly.
After the encryption process, a secret key will be generated.
Only the authorized user can use the secret key to decrypt the model, while without the secret key, the model will output wrong predictions.

The main contributions of this work are three-folds:
\begin{itemize}
  \item
  This paper proposes an active DNN IP protection method based on adversarial perturbation, which can prevent the infringement of DNN IP in advance.
  In order to achieve the function of active authorization control, the parameters that have the greatest impact on the performance of the model are selected by utilizing the gradient of the model's loss function, and the adversarial perturbations will be added to these selected parameters.
  With an extremely low number of encrypted parameters, the model's accuracy will be significantly decreased.
  After the encryption process, a secret key containing the positions of the encrypted parameters and the values of the added perturbations will be generated.
  With the secret key, the authorized user can decrypt the model and obtain high accuracy.

  \item
  Compared with the existing watermarking based verification methods, the proposed method can prevent unauthorized usage of the model, thus protects the DNN IP from infringement in advance.
  Besides, This method does not require to retrain the model.
  Compared with most of the existing active DNN IP protection works, which require to retrain the model, the proposed method is low-cost and more practical in the commercial applications.
  Moreover, the proposed method only needs to encrypt an extremely low number of parameters, which significantly reduces the storage overhead of the secret keys and the decryption time.

  \item
  Experimental results demonstrate the effectiveness of the proposed method, as the test accuracy of the encrypted model are only 10.36\%, 10.86\% and 6.94\% on Fashion-MNIST \cite{abs-1708-07747}, CIFAR-10 \cite{krizhevsky2009learning} and GTSRB \cite{stallkamp2011german}, respectively.
  The number of the encrypted parameters is as low as 23, and the proportion of the encrypted parameters is as low as 0.000205\% of all the parameters of the model.
  Besides, for malicious attackers, it's hard to detect the encrypted parameters, as the encrypted weights are all in a normal range and the change of the distribution of the weights is insignificant.
  Further, the robustness of the proposed method against three attacks, i.e., model fine-tuning attack, model pruning attack, and adaptive attack are experimentally demonstrated.
  Experimental results show that, the model's test accuracy is still as low as 17.64\% after the three attacks.

\end{itemize}

This paper is organized as follows.
The related works are introduced in Section \ref{relatedwork}.
The proposed adversarial perturbation based method is elaborated in Section \ref{proposedmethod}.
The experimental results are presented and analyzed in Section \ref{experiments}.
This paper is concluded in Section \ref{Conclusion}.

\section{Related Work} \label{relatedwork}
In this section, the related works including the watermarking methods and the active DNN IP protection methods are briefly reviewed, and the comparison between the proposed method and the related DNN IP protection works is also presented.

\textbf{Watermarking methods.}
Uchida \textit{et al.} \cite{UchidaNSS17} embed watermark into the parameters of a DNN model. The watermark can be extracted to verify the ownership of the model.
This method can only be applied in the while-box scenario, in which the model's parameters are publicly accessible.
However, in the commercial applications, DNN models are deployed in the black-box scenario in most cases, where the verifier have no access to the internal parameters of the pirated model.

Rouhani \textit{et al.} \cite{RouhaniCK19} proposed an IP protection framework which can work in black-box scenario.
They embed the watermark into the output layer of DNN model, and selects some input samples as the input key set.
With the presence of the input key set, the watermark can be extracted.
Chen \textit{et al.} \cite{DBLP:journals/corr/abs-1904-00344} embed the owner's signature to the model.
With the key images, the owner can decode his signature from the model and claim his ownership of the model.
Merrer \textit{et al.} \cite{DBLP:journals/nca/MerrerPT20} slightly modify the model's decision boundary to mark the model, which will output specific predictions when the adversarial examples are presented.
Thus, the owner can use adversarial examples to query the marked model and verify the ownership.

\textbf{Active DNN IP protection methods.}
Chen and Wu \cite{ChenW18} realize the function of access control by training an anti-piracy model, which will be dysfunctional for unauthorized users.
With the transformation module, users can generate the authorized input so as to normally use the model.
Pyone \textit{et al.} \cite{pyone2020training} shuffle the pixels of the training images, and use these training images to train the model.
The model only works when the input samples are preprocessed with the secret key.
Fan \textit{et al.} \cite{FanNC19} design some passports and embed them into the model.
Without these passports, the model will output wrong prediction results, thus prevents the unauthorized usage of the model.
Some DNN IP protection works are based on the hardware.
Chen \textit{et al.} \cite{ChenFRZK19} proposed an on-device IP protection method, named DeepAttest.
In DeepAttest, the pre-defined fingerprint will be encoded in the target DNN and later be extracted with the support of hardware platform to verify the legitimacy of the DNN program.
Only authorized DNN program is allowed to run on the target hardware device \cite{ChenFRZK19}.
Chakraborty \textit{et al.} \cite{chakraborty2020hardware} changes the loss function in the training process, which will make the trained model obfuscated.
They then embed the key into the target hardware device, without which the obfuscated model can not work properly.
Lin \textit{et al.} \cite{lin2020chaotic} encrypt the model by changing the position of the model's parameters.
After purchasing the secret key, the authorized user can decrypt the model, and make the model work properly.

The comparison between the proposed method and the existing DNN IP protection works, including the watermarking methods and the active authorization control methods are presented as follows.

\textbf{Comparison with the watermarking methods.}
The proposed method achieves the function of active authorization control where only authorized users are allowed to use the model.
Compared with the existing watermarking methods, which only work after the infringement occurs, the proposed method can prevent unauthorized users from using the model, thus protect the DNN IP in advance.
Moreover, most of the watermarking methods retrain the model to embed the watermark, which changes the parameters completely.
In contrast, the implementation of the proposed method only needs to encrypt an extremely small amount of parameters.

\begin{figure*}[ht]
\centerline{\includegraphics[width=4.5in]{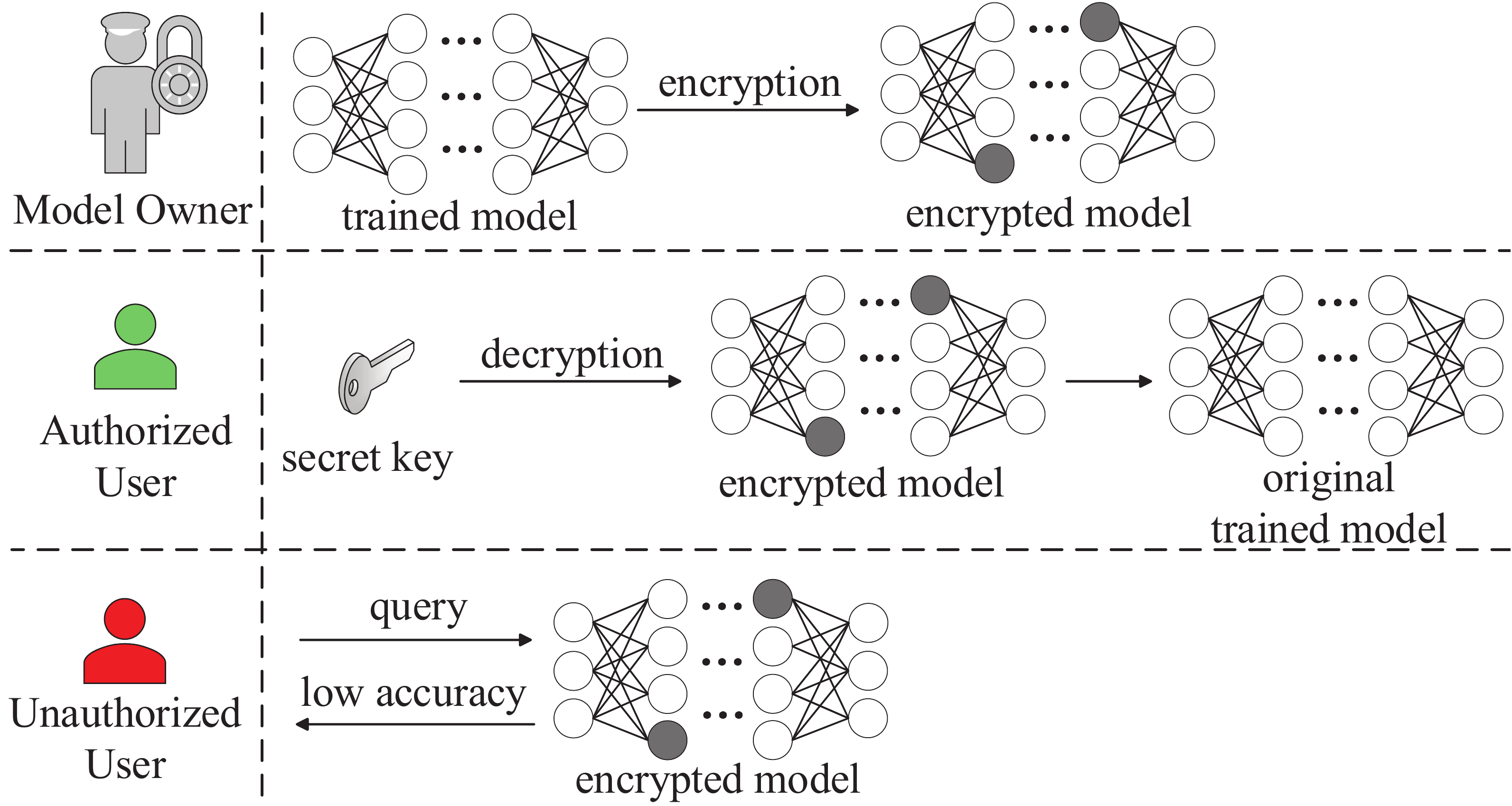}}
\caption{The overall flow of the proposed method.}
\label{fig1}
\end{figure*}

\textbf{Comparison with the existing authorization control methods.}
The proposed method encrypts the parameters of a trained model with well-crafted adversarial perturbations.
Compared with the works \cite{pyone2020training}, \cite{ChenW18}, \cite{FanNC19}, \cite{ChenFRZK19}, \cite{chakraborty2020hardware}, which require retraining the model \cite{pyone2020training}, \cite{ChenW18}, \cite{FanNC19}, \cite{chakraborty2020hardware}, or need the support of expensive hardware devices \cite{ChenFRZK19}, \cite{chakraborty2020hardware}, the proposed method requires low computational resources and does not need the hardware support, thus is more feasible in the realistic commercial applications.
Moreover, compared with \cite{lin2020chaotic}, which significantly modifies the model parameters by exchanging the positions of a substantial amount of parameters, the proposed method only needs to encrypt an extremely low number of parameters, which makes these encrypted parameters undetectable.

\section{The Proposed Method}\label{proposedmethod}

\subsection{Overall Flow} \label{overrallflow}
The overall flow of the proposed method is illustrated in Fig.~\ref{fig1}.
The proposed method includes the following three parts:
\begin{itemize}
  \item \textbf{Parameter encryption.}
  The process of parameter encryption includes three steps:
  First, a small amount of labeled images are sampled from the training dataset of the DNN model. The number of these sampled images is denoted as $N$.
  The set of the sampled images is referred as the encryption set.
  Second, the model owner randomly selects some convolutional and fully-connected layers from the model.
  With the encryption set, the adversarial perturbation is iteratively added on the weights of the selected layers, where in each iteration, there is only one hidden layer that will be selected for encryption, and only one weight of the selected layer will be perturbed with adversarial perturbation.
  For each selected layer, if the maximum number of iteration has been reached, this layer will not be encrypted any more, and another layer will be selected for encryption.
  Third, in each iteration, the encryption set will be input into the model, and the value of the model's loss function will be calculated, if the value of loss function is higher than the pre-defined encryption threshold, the encryption process will be terminated, which means the model has been successfully encrypted.

  \item \textbf{Secret key generation.}
  In the encryption process, the positions of the encrypted parameters and the values of the added adversarial perturbations will be stored.
  When the encryption process is terminated, the stored information will form the secret key.

  \item \textbf{Decryption.}
  With the secret key, the authorized user can remove the perturbation added to the model's parameters and restore the model's original high accuracy.
  Note that, the proposed method only needs to encrypt an small number of model's parameters (as low as 23, which will be discussed in Section \ref{Experimental Results}), thus the decryption time can be effectively reduced.

\end{itemize}
In the subsequent sections, the above three parts will be elaborated.

\subsection{Parameter Encryption} \label{Encryption}
The encryption process of the proposed method is inspired by JSMA \cite{papernot2016limitations}, which generates adversarial examples by perturbing only a few pixels of the clean images.
The process of the adversarial perturbation based parameter encryption is summarized in the proposed Algorithm \ref{alg1}.
First, the encryption set containing a small amount of labeled images is sampled from the model's training set.
Note that, the encryption set can also be sampled from the dataset that has the same distribution as that of the model's training set.
The encryption set can be denoted as $D_e = (X_e, Y_e)$, where $X_e = \{x_1,...,x_N\}, Y_e = \{y_1,...,y_N\}$ are the sampled images and the corresponding labels, respectively, and $N$ is the number of the sampled images.
Second, for a trained model, the model owner randomly selects some convolutional and fully-connected layers for encryption.
These randomly selected layers are referred as the encrypted layers.
Then, the adversarial perturbations are iteratively added to the weights of these encrypted layers.
In order to decrease the number of encrypted parameters, in each iteration, only one encrypted layer will be selected for encryption, and only one weight in the layer will be perturbed.
Formally, let $\cal{L}$ denotes the encrypted layer set.

Here, for a single layer $l \in {\cal{L}}$, we elaborate the following two issues: (i) how to select the weights that has significant impact on the performance of the model; (ii) how to perturb the selected wight.

\textbf{How to select the weight.}
Formally, let $F$ and $\tilde F$ denotes the original trained model and the encrypted model, respectively.
For an input image $x$, the predicted label of $F$ and $\tilde F$ are denotes as $F(x)$ and $\tilde F(x)$, respectively.
The loss function is denoted as $L$.
For the encryption set $D_e = (X_e, Y_e)$, the value of the loss function can be denoted as $L(F,{D_e})$.
The parameter weights of the layer $l$ can be denoted as ${W_l} = [{w_{l1}},{w_{l2}},...]$.
In each iteration, the gradient ${\nabla _{{W_l}}}L(F,{D_e})$ for the weights ${W_l}$ will be calculated as:
\begin{equation}
{\nabla _{{W_l}}}L(F,{D_e}) = [\frac{{\partial L(F,{D_e})}}{{\partial {w_{l1}}}},\frac{{\partial L( F,{D_e})}}{{\partial {w_{l2}}}},...]
\end{equation}
The components (partial derivatives) contained in the gradient ${\nabla _{{W_l}}}L(F,{D_e})$ reflect the descent rate of the loss function value $L(F,{D_e})$.
Thus, the partial derivative with the highest absolute value will be selected, and the corresponding weight $w_{lt}$ is considered to have the greatest impact on the model's performance among all the weights contained in layer $l$, i.e.,
\begin{equation}
\left| {\frac{{\partial L(F,{D_e})}}{{\partial {w_{lt}}}}} \right| = \max [\left| {\frac{{\partial L(F,{D_e})}}{{\partial {w_{l1}}}}} \right|,\left| {\frac{{\partial L( F,{D_e})}}{{\partial {w_{l2}}}}} \right|,...]
\end{equation}
then the weight $w_{lt}$ will be selected, and perturbed with adversarial perturbation.
The model's performance will be significantly deteriorated with the modification of weight $w_{lt}$.

\textbf{How to perturb the selected weight.}
After the weight $w_{lt}$ with the highest absolute value of partial derivative is selected, the value of the adversarial perturbation ${\eta _{lt}}$ can be calculated as:
\begin{equation}
{\eta _{lt}} = \theta  \times sign[\frac{{\partial L(F,{D_e})}}{{\partial {w_{lt}}}}] \times [\max({W_{l}}) - \min({W_{l}})]
\label{eq4}
\end{equation}
where $\theta$ is a hyper-parameter with low value that controls the value of the perturbation ${\eta _{lt}}$, and $\max({W_{l}})$ and $\min({W_{l}})$ are the maximum weight and the minimum weight in the layer $l$, respectively.
The weight $w_{lt}$ will be updated as $Clip({w_{lt}} + {\eta _{lt}})$.
For other weights that are not be selected, these weights will not be updated and remain the same.
The $Clip$ function is used to control the value of the updated weights, which can be formalized as:
\begin{equation}
\begin{array}{l}
{{Clip}}({w_{lt}}) = \left\{ {\begin{array}{*{20}{l}}
{{w_{lt}},{w_{lt}} \in [{T_{l1}},{T_{l2}}]}\\
{{T_{l1}},{w_{lt}} < {T_{l1}}}\\
{{T_{l2}},{w_{lt}} > {T_{l2}}}
\end{array}} \right.\\
\begin{array}{*{20}{c}}
{{T_{l1}} = \min ({W_l}) + \alpha  \times [\max ({W_l}) - \min ({W_l})]}\\
{{T_{l2}} = \max ({W_l}) - \alpha  \times [\max ({W_l}) - \min ({W_l})]}
\end{array}
\end{array}
\end{equation}
where $\alpha$ is a hyper-parameter, and $T_{l1}$ and $T_{l2}$ are two thresholds that control the value of the updated weights.
If the updated weight ${{w}_{lt}}+ {\eta _{lt}}$ excesses the range $[T_{l1}, T_{l2}]$ (lower than $T_{l1}$, or higher than $T_{l2}$), it will be clipped into the range $[T_{l1}, T_{l2}]$.
With the $Cilp$ function, the encrypted weights will be controlled into a normal range, thus makes these encrypted weights hard to be detected by potential attackers.
Further, if ${{w}_{lt}}+ {\eta _{lt}} \notin [T_{l1}, T_{l2}]$, after the weight ${{w}_{lt}}$ is updated as $Clip({w_{lt}} + {\eta _{lt}})$, then in the following iterations, the adversarial perturbations will not be added to the weight ${{w}_{lt}}$.

In each iteration, after the weights is added with adversarial perturbations, the encryption set $D_e$ will be input into the model, and the value of the loss function $L(F,{D_e})$ will be calculated.
If $L(F,{D_e})>T_{loss}$, the encryption process will be terminated, where $T_{loss}$ is the pre-defined encryption threshold.
Otherwise, the encryption process will continue until $L(F,{D_e})>T_{loss}$.
Note that, the encryption process will be terminated after a few iterations, as the modification of each selected parameter leads to a significant deterioration in the accuracy of the model, which makes the value $L(F,{D_e})$ of the loss function rapidly increase and higher than the threshold $T_{loss}$

\renewcommand{\algorithmicrequire}{\textbf{Iutput:}}
\renewcommand{\algorithmicensure}{\textbf{Output:}}
\begin{algorithm}[!htbp]
  \caption{Parameter encryption.}
  \label{alg1}
  \begin{algorithmic}[1]
    \Require
        encrypted layer set $\cal{L}$ and the corresponding weights $W_{l} = [{w_{l1}},{w_{l2}},...], (l \in {\cal{L}})$, trained model $F$, encryption set $D_e$, the maximum number of the iteration $I$, encryption threshold $T_{loss}$.
    \Ensure encrypted model $\tilde F$.
    \For {$l$ in $\cal{L}$}
        \State $Mask \leftarrow [1,1,...,1]$, where the dimension of $Mask$
        \Statex \quad \, is the same as that of $W_l$;
        \For {$i=1$ to $I$}
            \State Calculate the value of the loss function $L(F,{D_e})$;
            \If {$L(F,{D_e}) > T_{loss}$}
            \State \Return $F$;
            \EndIf
            \State Calculate the gradient ${\nabla _{{W_l}}}L(F,{D_e})$;
            \State ${\nabla _{{W_l}}}L(F,{D_e}) \leftarrow {\nabla _{{W_l}}}L(F,{D_e}) \odot Mask$;
            \State Select the the weight $w_{lt}$ with the highest
            \Statex \quad \, \, \, \, absolute value of partial derivative;
            \State ${\eta _{lt}} \leftarrow \theta  \times sign[\frac{{\partial L(F,{D_e})}}{{\partial {w_{lt}}}}] \times [\max({W_{l}}) - \max({W_{l}})]$;
            \If {${w_{lt}} + {\eta _{lt}} \notin [T_{l1}, T_{l2}]$}
                \State $Mask_t \leftarrow 0$;
            \EndIf
            \State ${w_{lt}} \leftarrow Clip({w_{lt}} + {\eta _{lt}})$;
        \EndFor

    \EndFor

    \State \Return $\tilde F$;
  \end{algorithmic}
\end{algorithm}

\subsection{Secret Key Generation} \label{keygeneration}
\begin{figure}[!htbp]
\centerline{\includegraphics[width=2.4in]{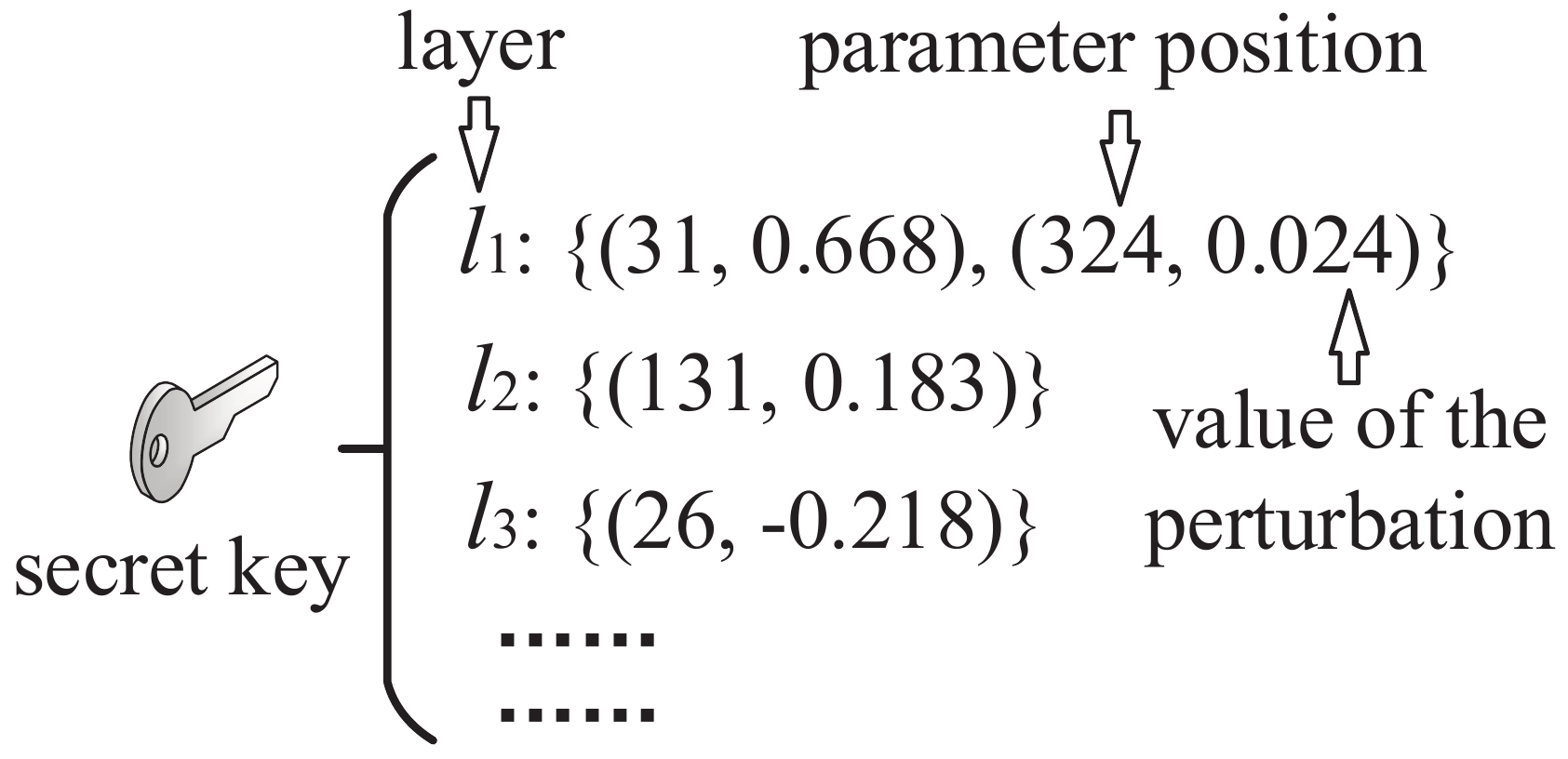}}
\captionsetup{font=normalsize}
\caption{An example of the secret key.}
\label{fig2}
\end{figure}

Fig. \ref{fig2} illustrates the composition of the secret key, which is generated after the parameter encryption.
Specifically, in each iteration, after the selected weight is updated, the position of the weight and the corresponding adversarial perturbation will be stored.
After the encryption, the stored information (positions and perturbations) will form the secret key $\cal{K}$, i.e.,
\begin{equation}
{\cal{K}} = \{ ({p_{lt}},{v_{lt}})|l \in {\cal{L}} ,t = 1,2,...n_l\}
\end{equation}
where ${\cal{L}}$ is the encrypted layer set, $l \in {\cal{L}}$ is the encrypted layer, $p_{lt}$ is the position of the encrypted parameter of the layer $l$, $v_{lt}$ is the added perturbation, and $n_l$ is the number of the encrypted parameters of layer $l$.
With the secret key, an authorized user can determine the position of the encrypted parameters for each layer, and remove the added adversarial perturbations to decrypt the model.
In addition, the number the encrypted parameters is as low as 23, and the proportion of the encrypted parameters is as low as 0.000205\% of all the model's parameters (as discussed in Section \ref{Experimental Results}).
Thus, in the proposed method, the storage overhead of the secret key is extremely low.

\subsection{Decryption} \label{Decryption}
The decryption process is described in the proposed Algorithm \ref{alg2}.
Assume there are $K$ layers $l_1,l_2,...,l_K$ contained in the encrypted model $\tilde F$.
For each layer $l \in \{l_1,l_2,...,l_K\}$, the weights of the layer $l$ is denoted as $\tilde W_l = [\tilde w_{l1}, \tilde w_{l2},...]$.
With the secret key ${\cal{K}} =\{({p_{lt}},{v_{lt}})|l \in {\cal{L}} ,t = 1,...,n_l\}$, the perturbations added to the encrypted model $\tilde F$ can be removed, and the original high accuracy of the trained model $F$ will be restored.
Specifically, for a layer $l \in \{l_1,l_2,...,l_K\}$, the positions of the encrypted parameters can be determined according to the secret key ${\cal{K}}$.
Let $p$ denotes the position of an encrypted parameter of the layer $l$, the added perturbation can be removed as:
\begin{equation}
w_{lp} = (\tilde w_{lp} - v_{lp})
\end{equation}
where $\tilde w_{lp}$ is the weight of the encrypted parameter, and $v_{lp}$ is the added perturbation for $\tilde w_{lp}$.

\renewcommand{\algorithmicrequire}{\textbf{Iutput:}}
\renewcommand{\algorithmicensure}{\textbf{Output:}}
\begin{algorithm}[!htbp]
  \caption{Decryption.}
  \label{alg2}
  \begin{algorithmic}[1]
    \Require
        secret key ${\cal{K}} =\{({p_{lt}},{v_{lt}})|l \in {\cal{L}} ,t = 1,...,n_l\}$, the encrypted model $\tilde F$ and it's weights $\tilde W_{l}$ of layer $l$, ($l \in \{l_1,l_2,...,l_K\}$).
    \Ensure original trained model $F$.
    \For {$i=1$ to $K$}
        \If {$l_i \in {\cal{L}}$}
        \State $l \leftarrow l_{i}$;
        \For {$j=1$ to $n_l$}

            \State $p \leftarrow p_{lj}$;
            \State $v \leftarrow v_{lj}$;
            \State $w_{lp} \leftarrow (\tilde w_{lp} - v)$;

        \EndFor
        \EndIf
    \EndFor

    \State \Return $F$;
  \end{algorithmic}
\end{algorithm}

\section{Experiments} \label{experiments}
In this section, the experiment is conducted to evaluate the proposed method.
First, the experimental setup is introduced in Section \ref{exp_setup}.
Then, the proposed method is evaluated and the experimental results are presented and analyzed in Section \ref{Experimental Results}.
Besides, the parameter discussion is presented in Section \ref{Parameter Discussion}.
Further, the robustness of the proposed method against three attacks is evaluated in Section \ref{Robustness}.
The proposed method is compared with the related works in Section \ref{Comparison}.

\subsection{Experimental Setup} \label{exp_setup}

\subsubsection{\textbf{Datasets}} \label{datasets}
The datasets used in the experiment are Fashion-MNIST \cite{abs-1708-07747}, CIFAR-10 \cite{krizhevsky2009learning} and GTSRB \cite{stallkamp2011german}.
For all the three datasets, the number of the encryption images contained in the encryption set $D_e$ is 300, i.e., $|D_e|=300$.
For Fashion-MNIST \cite{abs-1708-07747}, CIFAR-10 \cite{krizhevsky2009learning}, and GTSRB \cite{stallkamp2011german}, the proportions of the encryption images of all the training images are only 0.5\% (300/60,000), 0.6\% (300/50,000), and 0.77\% (300/39,208), respectively.
The loss functions used in the three datasets are all cross-entropy loss, and the encryption threshold $T_{loss}$ is set to be 15, 12 and 12 for Fashion-MNIST \cite{abs-1708-07747}, CIFAR-10 \cite{krizhevsky2009learning} and GTSRB \cite{stallkamp2011german}, respectively.
The hyper-parameter $\theta$ is set to be 0.07 for all the three datasets, and the hyper-parameter $\alpha$ is set to be 0.05, 0.05 and 0.03 for Fashion-MNIST \cite{abs-1708-07747}, CIFAR-10 \cite{krizhevsky2009learning} and GTSRB \cite{stallkamp2011german}, respectively.

\subsubsection{\textbf{DNN models}}
The model trained on the Fashion-MNIST \cite{abs-1708-07747} dataset is the DenseNet model \cite{DBLP:conf/cvpr/HuangLMW17}.
The DenseNet model is trained on the training set of Fashion-MNIST for 20 epochs with the SGD optimizer \cite{Theodoridis2015Stochastic}.
The learning rate and batch size are set to be 0.1 and 128, respectively.
The DenseNet model contains 46 convolutional or fully-connected layers.
The maximum number of the iteration $I$ is set to be 18 for the DenseNet model.
In the experiment, only 6 layers of DenseNet model are selected for encryption by Algorithm \ref{alg1}.

The model trained on the CIFAR-10 \cite{krizhevsky2009learning} dataset is the ResNet model \cite{HeZRS16}.
The ResNet model is trained on the training set of CIFAR-10 for 200 epochs with the SGD optimizer \cite{Theodoridis2015Stochastic}.
The batch size is set to be 128.
Initially, the learning rate for ResNet model is set to be 0.1, and the \textit{CosineAnnealingLR} function \cite{DBLP:conf/iclr/LoshchilovH17} is used to decrease the learning rate in the training process.
The ResNet model contains 58 convolutional or fully-connected layers.
The maximum number of the iteration $I$ is set to be 18 for the ResNet model.
In the experiment, only 14 layers of ResNet model are selected for encryption by the proposed Algorithm \ref{alg1}.

The model trained on the GTSRB \cite{stallkamp2011german} dataset is the AlexNet model \cite{KrizhevskySH17}.
The AlexNet model is trained on the training set of GTSRB for 200 epochs with the Adam optimizer \cite{DBLP:journals/corr/KingmaB14}.
The learning rate and batch size are set to be 0.003 and 128, respectively.
The ResNet model contains 8 convolutional or fully-connected layers.
The maximum number of the iteration $I$ is set to be 108 for the AlexNet model.
Note that, the number $I$ for the AlexNet model is much higher than that for the DenseNet model or ResNet model, this is because the number of layers contained in AlexNet is much lower than that of DenseNet and ResNet.
In the experiment, only 4 layers of AlexNet model are selected for encryption by Algorithm \ref{alg1}.

\subsubsection{\textbf{Metircs}} \label{metrics}
\textbf{accuracy drop \bm{$A_{d}$}.}
This metric represents the drop of the model's accuracy after parameter encryption.
Let $A_o$ represent the original accuracy of the trained model, and $A_e$ represents the accuracy of the encrypted model.
Then, the accuracy drop is $A_{d} = A_o - A_e$.
A large value of $A_{d}$ indicates that the proposed method can effectively deteriorate the model's performance.
The larger the value of $A_{d}$, the more effective the proposed method is.

\subsection{\textbf{Experimental Results}} \label{Experimental Results}
The drop of the model's accuracy caused by parameter encryption is presented in Table \ref{tab1}.
The values of accuracy drop $A_d$ are as high as 80.65\%, 81.16\% and 87.91\% for Fashion-MNIST \cite{abs-1708-07747}, CIFAR-10 \cite{krizhevsky2009learning} and GTSRB \cite{stallkamp2011german}, respectively.
For GTSRB, the accuracy of the model is reduced from 94.85\% to 6.94\%, which indicates that the model's performance has been significantly degraded.
Besides, for Fashion-MNIST and CIFAR-10, the proposed method can reduce the the model's accuracy to around 10\%.
As both the two datasets contain 10 classes, the model's accuracy is reduced to a random guess (i.e., close to 10\%).
In conclusion, the above experimental results demonstrate the effectiveness of the proposed method.

\begin{table}[!htbp]
\caption{The accuracy drop $A_{d}$ of the model after parameter encryption. The larger the value of $A_{d}$, the more effective the proposed method is.}
  \begin{center}
    \begin{tabular}{|c|c|c|c|}
    \hline
    \multirowcell{3}{Dataset} &\multicolumn{2}{c|}{Test accuracy} & \multirowcell{3}{Accuracy\\ drop $A_{d}$} \\
    \cline{2-3}
    \multicolumn{1}{|c|}{} & \makecell[c]{Before\\ encryption} & \makecell[c]{After\\ encryption} & \\
    \hline
    \makecell[c]{Fashion-MNIST (DenseNet)} & 91.01\% & 10.36\% & 80.65\% \\
    \hline
    \makecell[c]{CIFAR-10 (ResNet)} & 92.02\% & 10.86\% & 81.16\% \\
    \hline
    \makecell[c]{GTSRB (AlexNet)} & 94.85\% & 6.94\% & 87.91\% \\
    \hline
    \end{tabular}
  \label{tab1}
  \end{center}
\end{table}

Let $n_e$ and $n_{all}$ denote the number of the encrypted parameters and the number of all parameters of the model, respectively.
For the three datasets, The proportion of the encrypted parameters of all the model's parameters (i.e., $n_e / n_{all}$) are presented in Table \ref{tab2}.
For the three datasets, the numbers of the encrypted parameters are all in an extremely low level (only 23, 47 and 48 for Fashion-MNIST \cite{abs-1708-07747}, CIFAR-10 \cite{krizhevsky2009learning} and GTSRB \cite{stallkamp2011german}, respectively).
Besides, the proportions of the encrypted parameters of all the model's parameters are 0.00698\%, 0.00552\% and 0.000205\% for Fashion-MNIST \cite{abs-1708-07747}, CIFAR-10 \cite{krizhevsky2009learning} and GTSRB \cite{stallkamp2011german}, respectively.
The reason is that, the proposed adversarial perturbation based method can search for a few parameters that have the most significant impact on the performance of the model.
Thus, with an extremely low number of encrypted parameters, the model's accuracy can be effectively decreased.

\renewcommand\arraystretch{1.2}
\begin{table}[!htbp]
  \centering
  \caption{The proportion of the encrypted parameters of all the model's parameters for the three datasets}
    \begin{tabular}{|c|c|c|c|c|}
    \hline
    \multicolumn{1}{|c|}{Dataset} & $n_e$ & $n_{all}$ & \makecell[c]{Proportion}  \\
    \hline
    Fashion-MNIST  & 23 & 329,677   & 0.00698\%  \\
    \hline
    CIFAR-10       & 47 & 850,864    & 0.00552\%  \\
    \hline
    GTSRB          & 48 & 23,398,080   & 0.000205\%  \\
    \hline
    \end{tabular}
  \label{tab2}
\end{table}

\begin{figure*}[!htbp]
\centerline{\includegraphics[width=7.2in]{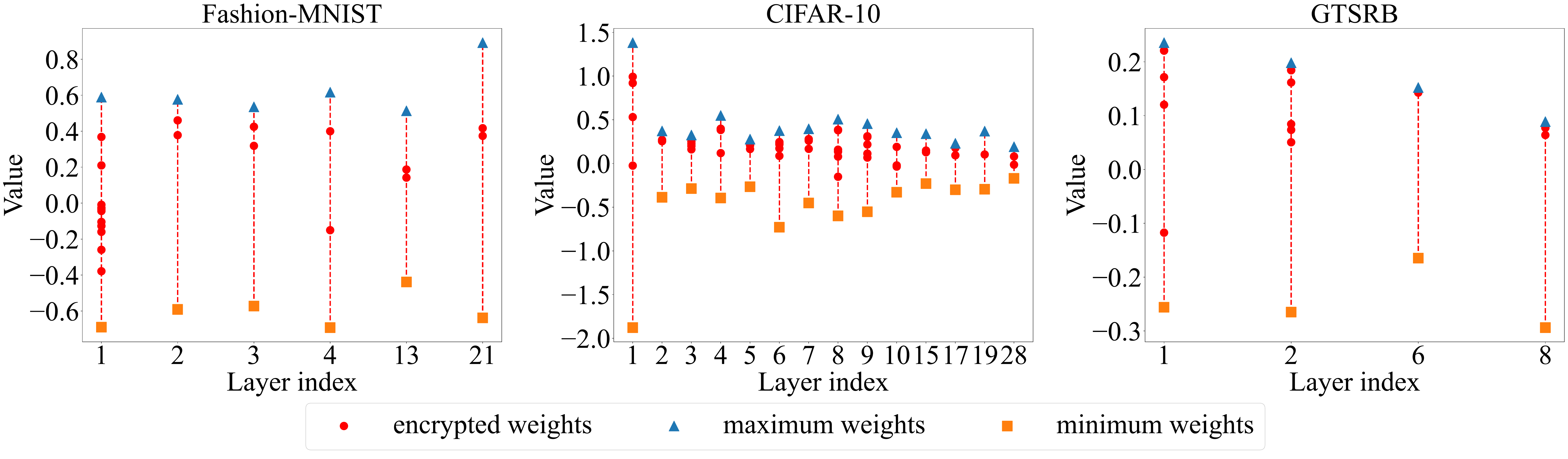}}
\captionsetup{font=normalsize}
\caption{The maximum and the minimum weights of the selected layer before encryption, and the weights of the encrypted parameters after encryption. Note that, some different encrypted parameters may have the same weights, thus some red dots (modified weights) may overlap (especially for GTSRB).}
\label{fig3}
\end{figure*}

Fig. \ref{fig3} depicts the maximum weights and the minimum weights of each middle layer before encryption, and the weights of the encrypted parameters of each middle layer after encryption.
As shown in Fig. \ref{fig3}, the numbers of the encrypted layers are only 6, 14, and 4 for Fashion-MNIST \cite{abs-1708-07747}, CIFAR-10 \cite{krizhevsky2009learning} and GTSRB \cite{stallkamp2011german}, respectively.
In addition, for each encrypted layer, the values of the modified weights are all in a normal range, i.e., higher than the minimum weights, and lower than the maximum weights of the original trained model, which effectively decreases the risk of being detected.
The reason is that, the proposed method utilize the $Clip$ function to control the values of the modified weights, which ensures that these encrypted weights all fall in the pre-defined range $[T_{l1},T_{l2}]$ (as discussed in Section \ref{Encryption}).

\begin{figure*}[!htbp]
\captionsetup[subfigure]{font=normalsize,margin=92pt}
\hspace{0in}
\subfloat[]{
\label{fig4a}
\includegraphics[width=0.32\linewidth]{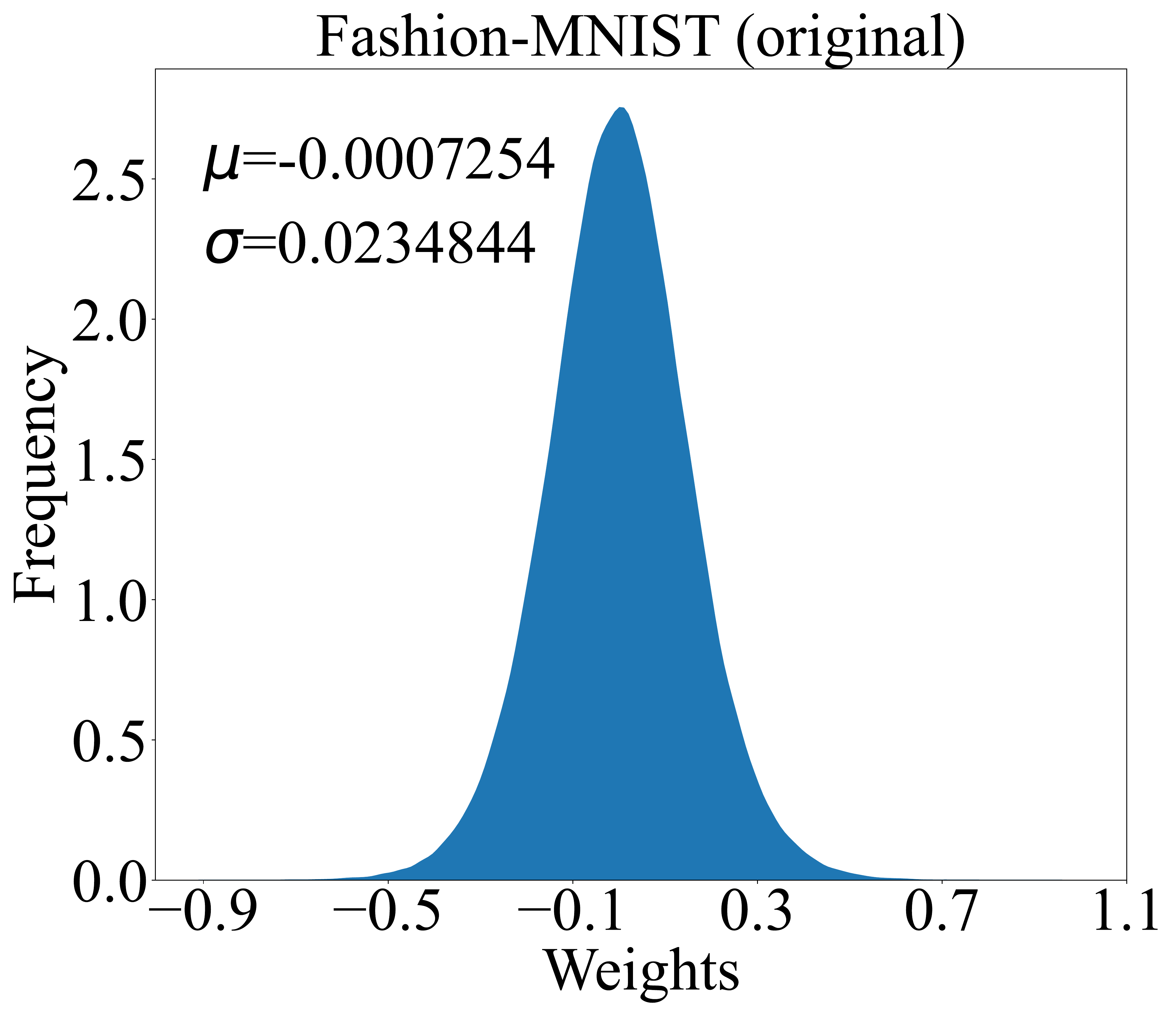}}
\hspace{0in}
\subfloat[]{
\label{fig4b}
\includegraphics[width=0.32\linewidth]{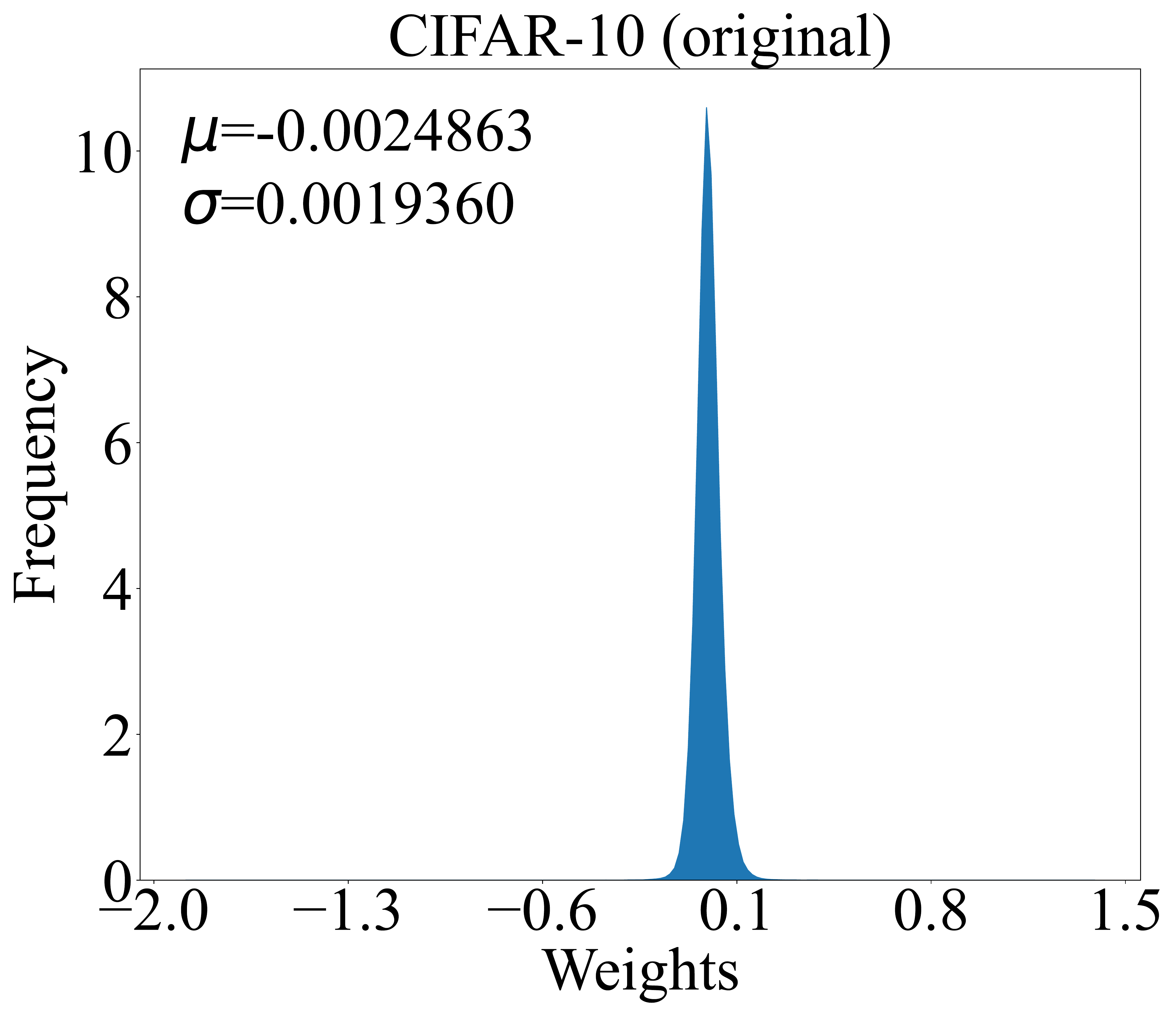}}
\hspace{0in}
\subfloat[]{
\label{fig4c}
\includegraphics[width=0.32\linewidth]{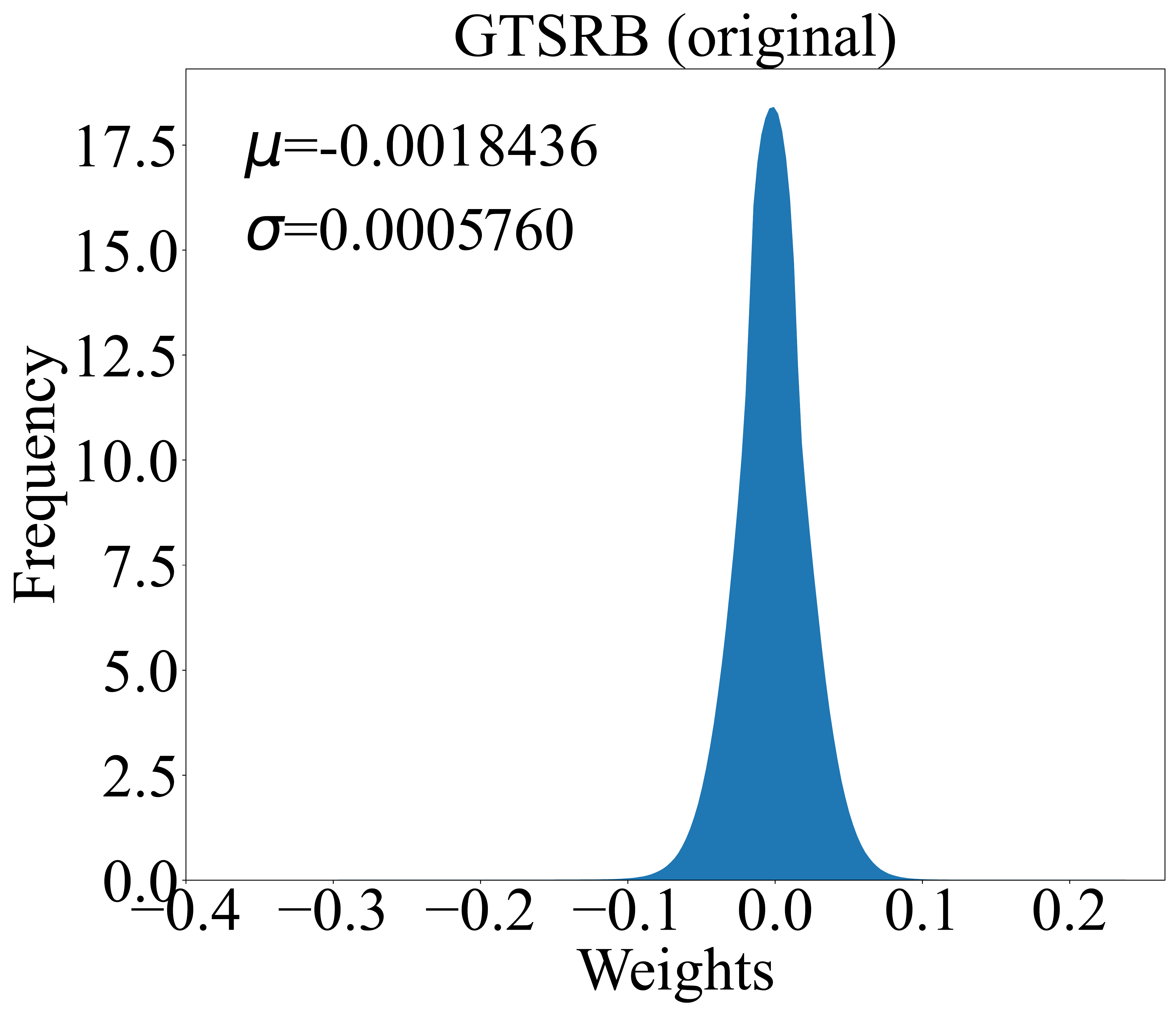}}

\hspace{0in}
\subfloat[]{
\label{fig4d}
\includegraphics[width=0.32\linewidth]{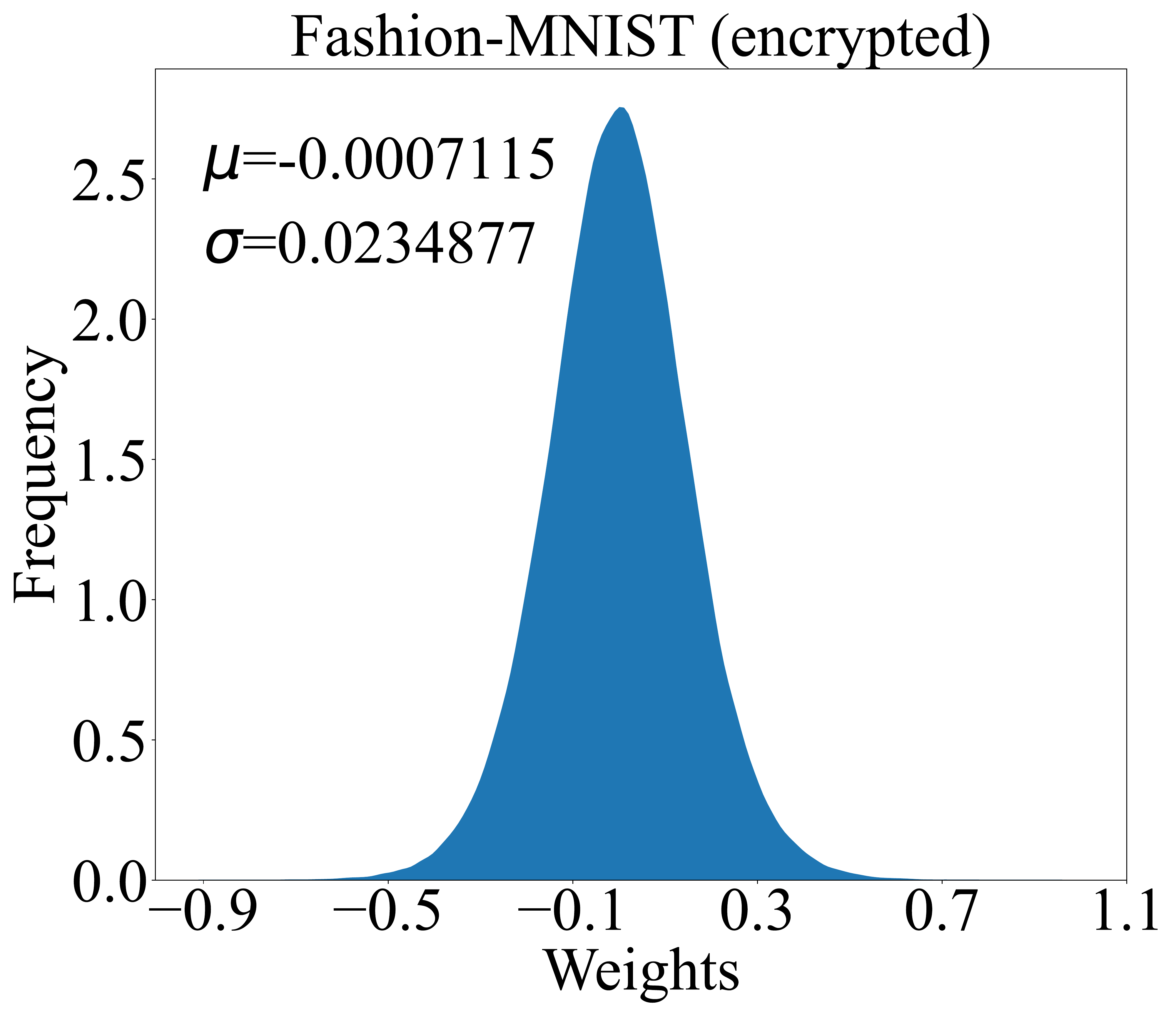}}
\hspace{0.01in}
\subfloat[]{
\label{fig4e}
\includegraphics[width=0.32\linewidth]{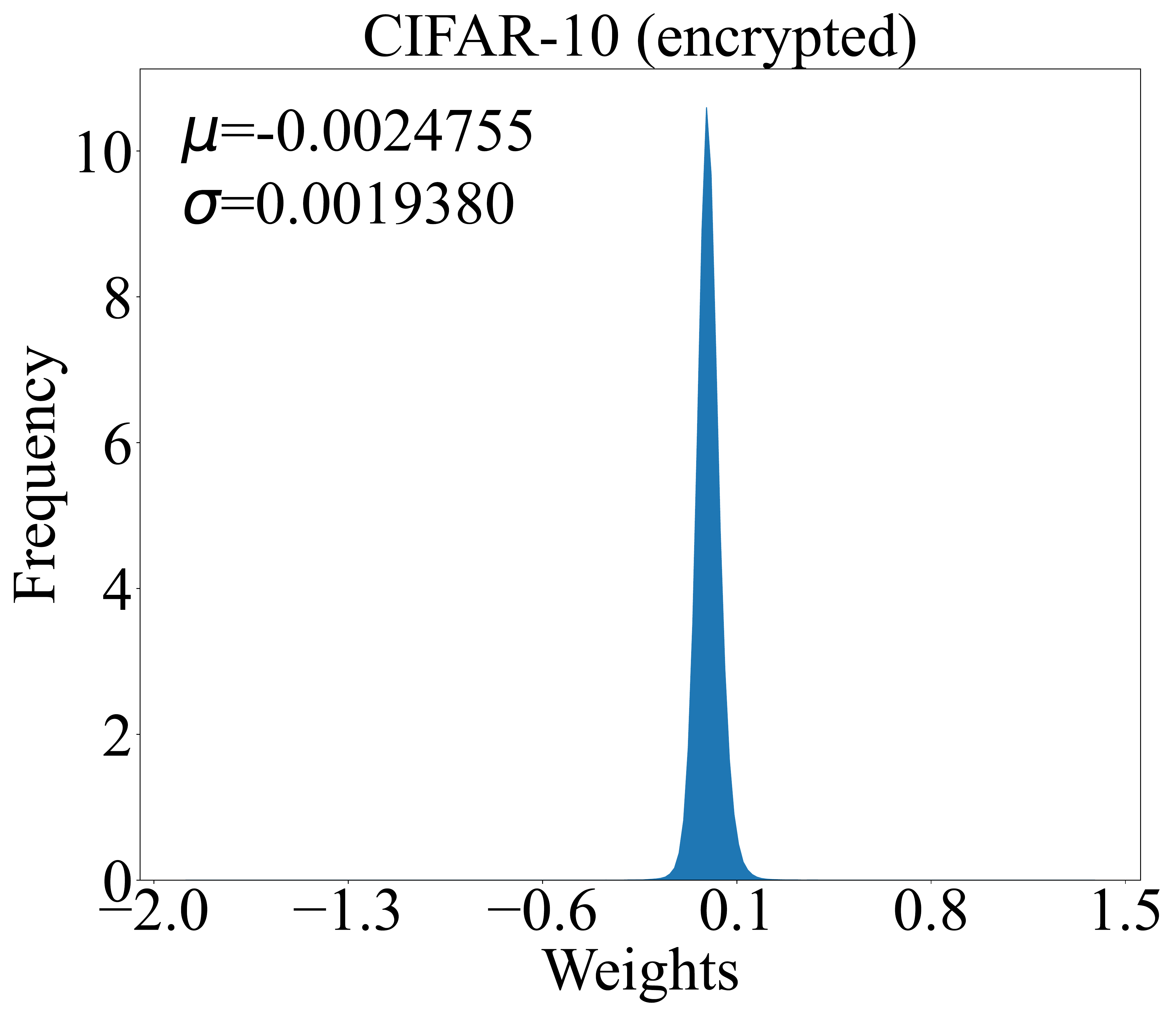}}
\hspace{0in}
\subfloat[]{
\label{fig4f}
\includegraphics[width=0.32\linewidth]{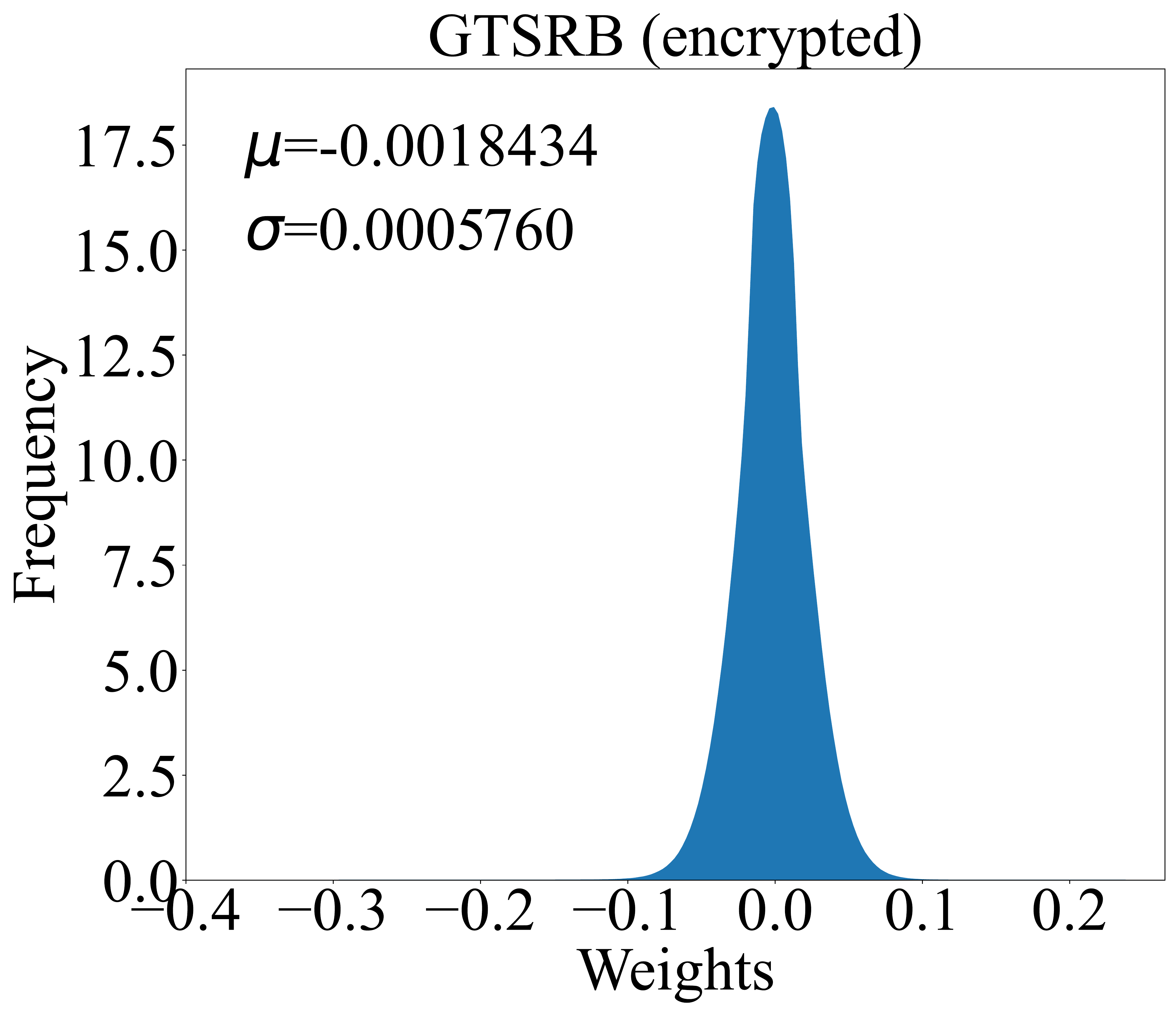}}
\hspace{0in}
\caption{The distribution of the weights of DNN model before and after encyption. Fig. \ref{fig4}(a)$\sim$(c) represent the distribution of weights of the original trained model for Fashion-MNIST, CIFAR-10 and GTSRB, respectively. Fig. \ref{fig4}(d)$\sim$(f) represent the distribution of weights of the encrypted model for Fashion-MNIST, CIFAR-10 and GTSRB, respectively. For all the three datasets, the mean value $\mu$ and variance value $\sigma$ of the distribution both remain the same after the parameter encryption.}
\label{fig4}
\end{figure*}

The distribution of the weights of DNN model before and after the parameter encryption is illustrated in Fig. \ref{fig4}.
As shown in Fig. \ref{fig4}, the change of the distribution of weights of DNN model is negligible.
The mean value $\mu$ and variance value $\sigma$ of the distribution of weights before and after the encryption is also calculated and presented in Fig. \ref{fig4}.
It can be seen that, for all the three datasets, the change of the mean value $\mu$ and variance value $\sigma$ of the distribution of weights is negligible after the encryption.
The above experimental results indicate that, the encrypted parameters are concealed for the potential attackers, as these encrypted parameters are extremely hard to be detected by analysing the distribution of weights.

\subsection{\textbf{Parameter Discussion}} \label{Parameter Discussion}
In this section, we evaluate and discuss the impact of three factors on the performance of the proposed method, which are the encryption threshold $T_{loss}$, the encrypted layers, and the number $n_e$ of the encrypted parameters.

\begin{figure}[!htbp]
\centerline{\includegraphics[width=2.2in]{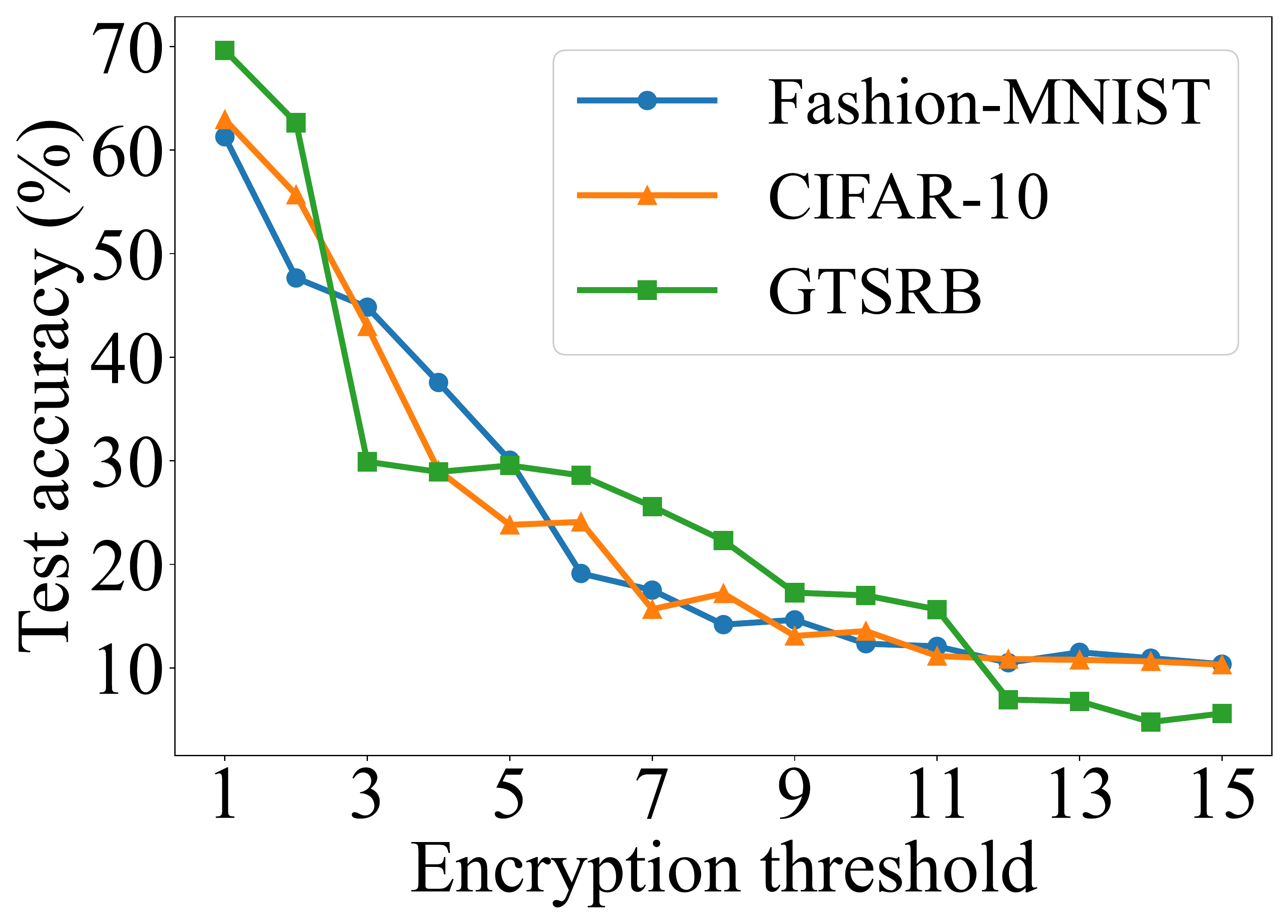}}
\captionsetup{font=normalsize}
\caption{The performance of the proposed method with different encryption threshold $T_{loss}$.}
\label{fig5}
\end{figure}

\begin{figure*}[!htbp]
\captionsetup[subfigure]{font=normalsize,margin=85pt}
\hspace{0in}
\subfloat[]{
\label{fig6a}
\includegraphics[width=0.3\linewidth]{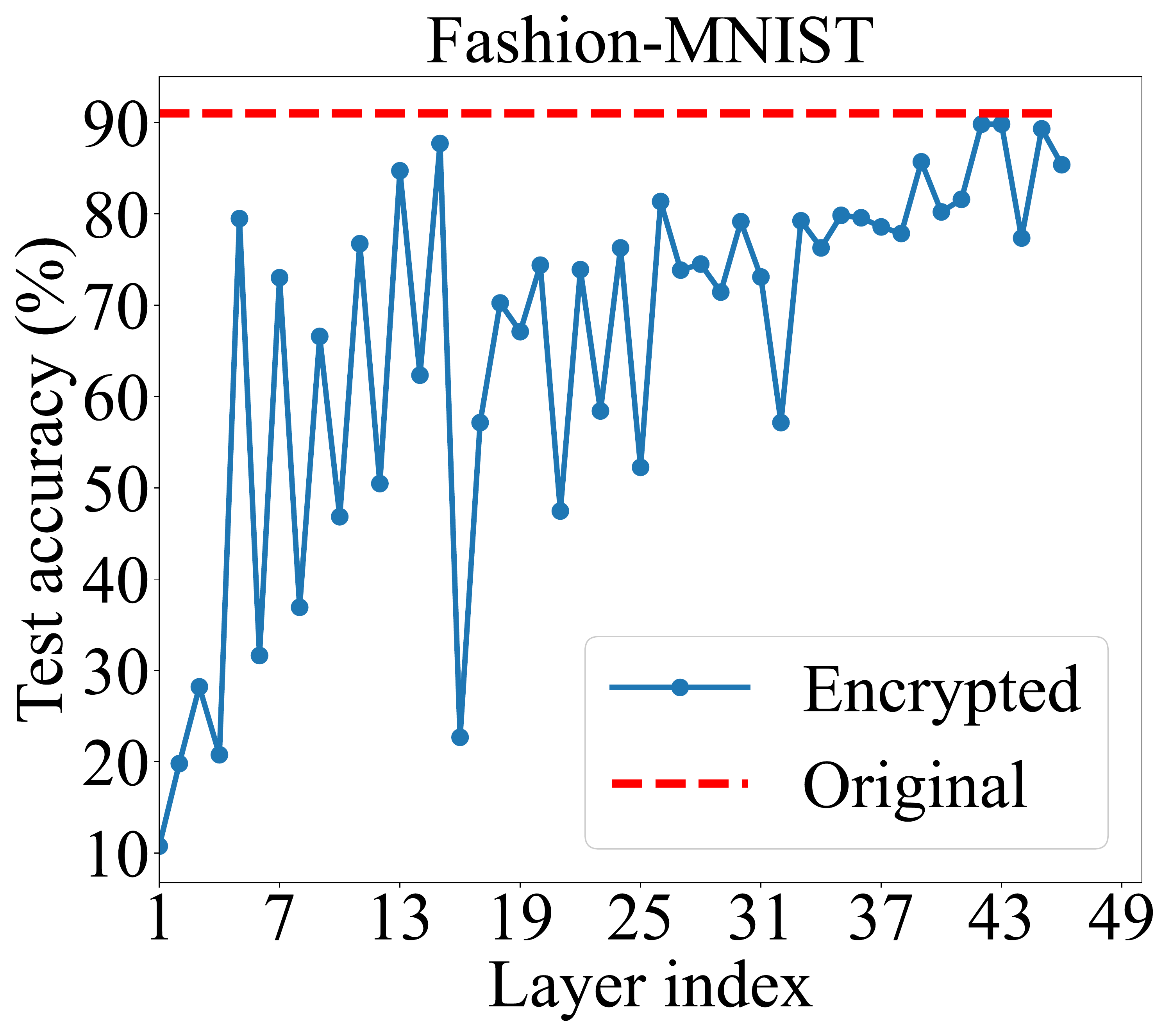}}
\hspace{0in}
\subfloat[]{
\label{fig6b}
\includegraphics[width=0.3\linewidth]{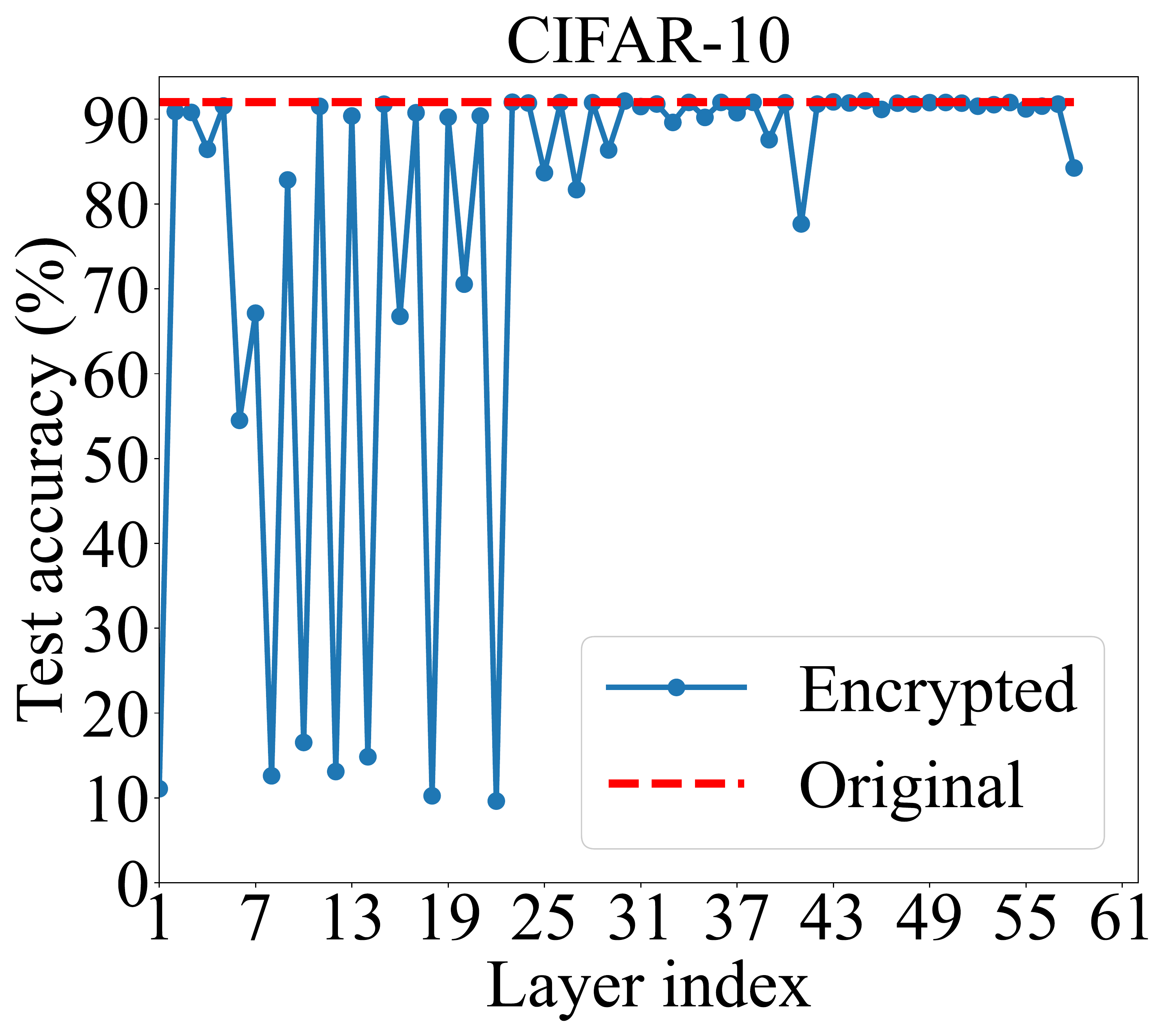}}
\subfloat[]{
\label{fig6b}
\includegraphics[width=0.3\linewidth]{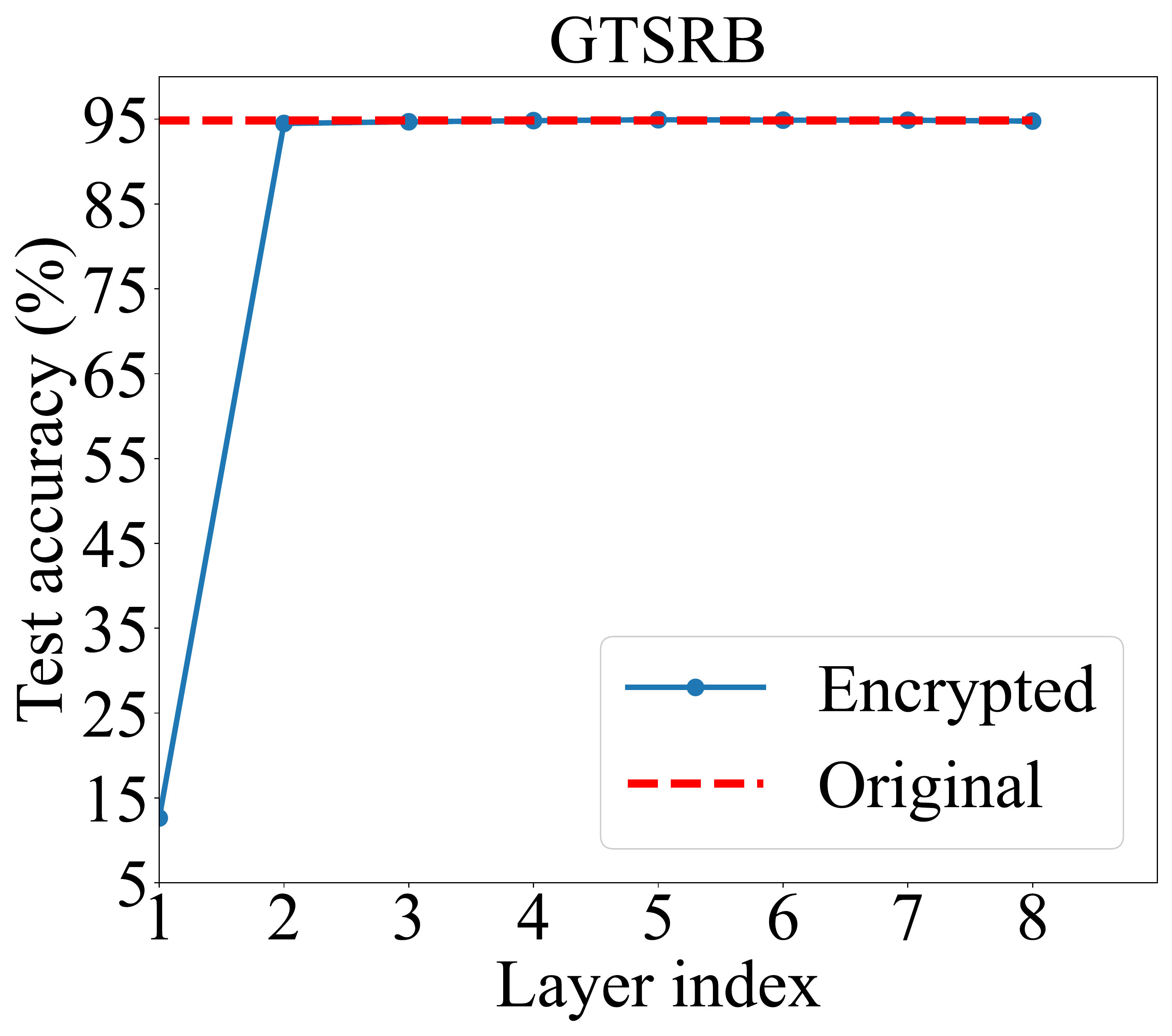}}
\caption{The test accuracy of DNN model with different encrypted layers, where there is only one layer that is selected for encryption in each experiment. Fig. \ref{fig6}(a)$\sim$(c) represent the experimental results for Fashion-MNIST, CIFAR-10 and GTSRB, respectively.}
\label{fig6}
\end{figure*}

\subsubsection{\textbf{Different Settings of Encryption Threshold \bm{$T_{loss}$}}} \label{Different Encryption Threshold}
The performance of the proposed method with different encryption threshold $T_{loss}$ is evaluated.
In the experiment, the values of $T_{loss}$ are set to be 1$\sim$15, and for each $T_{loss}$, the test accuracy of the encrypted model is calculated, as shown in Fig. \ref{fig5}.
When $T_{loss}=1$, the test accuracies are 61.28\%, 62.95\%, and 69.34\% on Fashion-MNIST, CIFAR-10, and GTSRB, respectively.
It can be seen that, the test accuracy of the encrypted model shows downward trends with the increase of the encryption threshold $T_{loss}$.
Besides, as shown in Fig. \ref{fig5}, when the encryption threshold $T_{loss}$ reaches 12, the test accuracy is reduced to a stable low value (around 10\%, 10\%, and 6\% on Fashion-MNIST, CIFAR-10, and GTSRB, respectively).
The experimental results indicate that, when $T_{loss}$ reaches 12, the test accuracy of the model can not be decreased by increasing the value of $T_{loss}$.
Moreover, the higher the value of $T_{loss}$, the more time it takes to encrypt the model.
Thus, $T_{loss}=$12 can be considered to have the optimal performance on the trade-off between the encryption time and decreasing the accuracy of the model.

\subsubsection{\textbf{Performance with Different Encrypted Layers}}
The performance of the proposed method with different encrypted layers is evaluated in this section.
In each experiment, there is only one layer that is selected for encryption, i.e., $|\cal{L}|=1$, where $\cal{L}$ is the encrypted layer set, and after the encryption, the test accuracy of the model is calculated.
The test accuracy of the encrypted model with different encrypted layers are presented in Fig. \ref{fig6}.
As shown in Fig. \ref{fig6}(a), for Fashion-MNIST, when only encrypt the layer 1, the test accuracy is effectively reduced from 91.01\% to 10.75\%.
Moreover, when only encrypting the layer 2, the layer 4, and the layer 16, the test accuracy of the encrypted model is as low as 19.79\%, 20.76\%, and 22.67\%, respectively.
As shown in Fig. \ref{fig6}(b), for CIFAR-10, the test accuracy is significantly decreased from 92.02\% to 9.63\% by encrypting the layer 22.
Moreover, when other layers are selected for encryption, the test accuracy can also be decreased to as low as around 10\%.
For GTSRB, when the layer 1 is encrypted, the test accuracy of DNN model is effectively reduced from 94.85\% to 12.64\%.
Besides, as shown in Fig. \ref{fig6}(c), when encrypting the layer 2$\sim$8, the test accuracy of the model is hardly decreased, which means the impact of the layer 2$\sim$8 on the performance of the model is insignificant for GTSRB.
In conclusion, the layer 1, the layer 22, and the layer 1 can be considered to have the optimal performance in terms of decreasing the accuracy of the model for Fashion-MNIST, CIFAR-10, and GTSRB, respectively.
Besides, the experimental results also indicate that, the proposed method can effectively decrease the model's accuracy by only encrypting a single layer of the model.

\subsubsection{\textbf{Different Settings of the Number of Encrypted Parameters}}
\begin{figure}[!htbp]
\centerline{\includegraphics[width=2.2in]{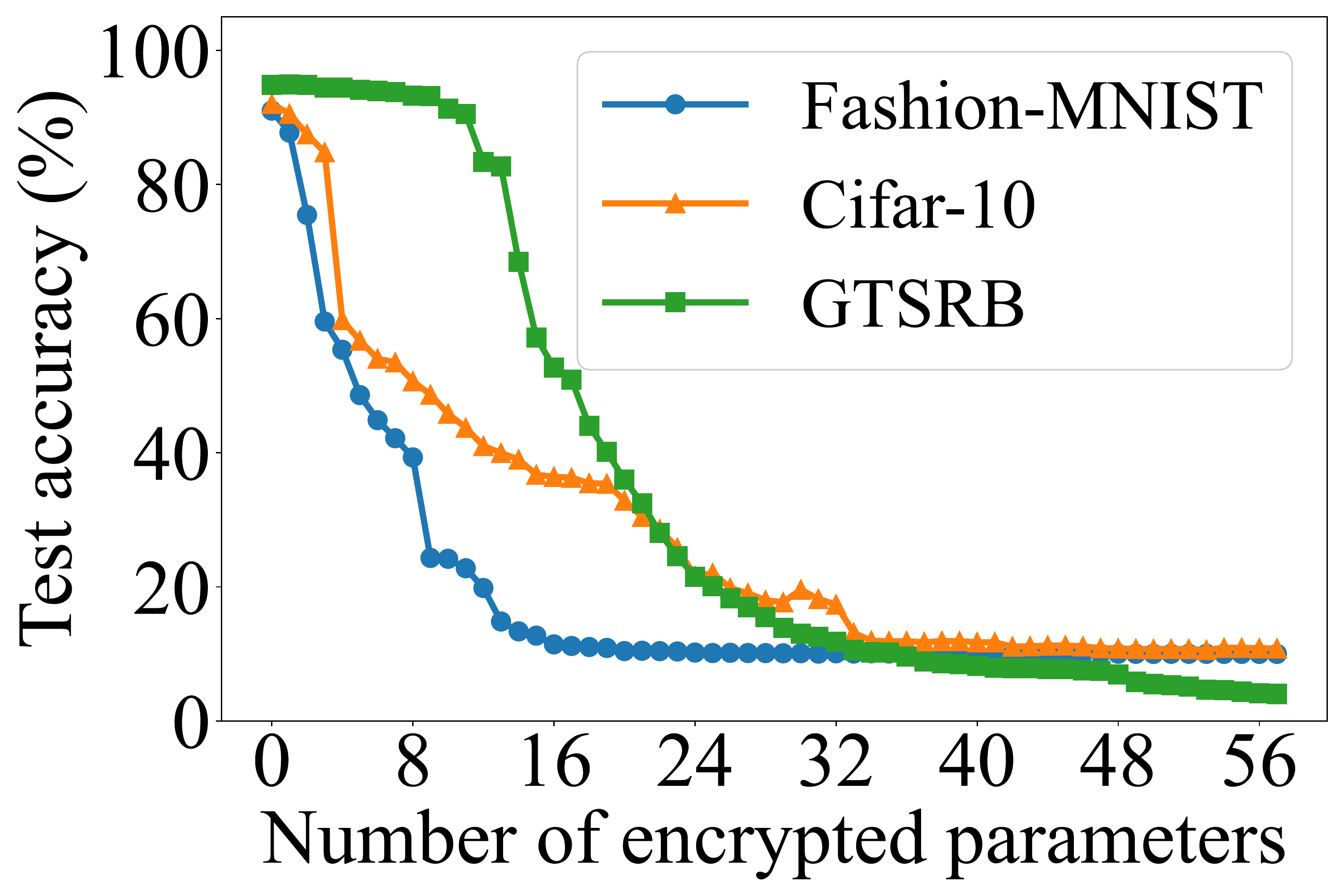}}
\captionsetup{font=normalsize}
\caption{The test accuracy of DNN model with different settings of the number of encrypted parameters.}
\label{fig7}
\end{figure}

The impact of different settings of the number of encrypted parameters on the performance of the proposed method is evaluated in this section.
With different settings of the number $n_e$ of encrypted parameters, the test accuracy of the encrypted model is depicted in Fig. \ref{fig7}.
As $n_e$ rises, the test accuracy shows downward trends for all the three datasets.
Moreover, for Fashion-MNIST and CIFAR-10, the test accuracy stabilized to a random guess (around 10\%) when $n_e$ reaches 16 and 34, respectively.
For GTSRB, when $n_e$ reaches 49, the test accuracy stabilized to an extremely low value (around 6\%).
It can be seen that, the test accuracy of the model can not be decreased by increasing $n_e$ when $n_e$ rises to 16, 34, and 49 for Fashion-MNIST, CIFAR-10, and GTSRB, respectively.
The above experimental results indicate that, by encrypting an extremely low number of encrypted parameters, the proposed method can achieve an excellent encryption effect.

\subsection{\textbf{Robustness of the Proposed Method}} \label{Robustness}
In this section, the robustness of the proposed method against three attacks are evaluated, which are model fine-tuning attack, model pruning attack, and adaptive attack.
In the adaptive attack, the attacker knows the detailed steps of the proposed method described in Section \ref{proposedmethod}.

\subsubsection{\textbf{Robustness against Model Fine-Tuning Attack}}
Model fine-tuning \cite{PittarasMMP17} is a method for transfer learning.
With model fine-tuning attack, an attacker has a chance to remove the perturbations of the encrypted parameters and restore the original accuracy of the model.
In the experiment, 10\% of the model's test set is used as the attacker's private dataset to fine-tune the encrypted model, and the accuracy of the fine-tuned model is evaluated with the rest 90\% of the test set.
The encrypted model is fine-tuned for 100 epochs and the test accuracy of the model is calculated for every 10 epochs.
For the three datasets, the test accuracy of the encrypted model after model fine-tuning attack is shown in Table \ref{tab3}.
The experimental results indicate that, the accuracy of the encrypted model remains at a low level (51.14\%, 43.80\%, and 61.66\% for Fashion-MNIST, CIFAR-10, and GTSRB, respectively) after fine-tuning attack even the number of epochs for fine-tuning attack reaches 100, which means that the proposed method is robust against model fine-tuning attack.

\renewcommand\arraystretch{1.2}
\begin{table}[!htbp]
  \centering
  \caption{The test accuracy of the encrypted model after model fine-tuning attack.}
    \begin{tabular}{|c|c|c|c|}
    \hline
    \multicolumn{1}{|c|}{Epochs} & Fashion-MNIST & CIFAR-10 & GTSRB \\
    \hline
    10      & 11.06\% & 30.69\% & 33.52\% \\
    \hline
    20      & 15.01\% & 35.43\% & 41.70\% \\
    \hline
    30      & 24.71\% & 37.77\% & 45.61\% \\
    \hline
    40      & 30.61\% & 37.99\% & 48.37\% \\
    \hline
    50      & 32.71\% & 39.12\% & 50.92\% \\
    \hline
    60      & 34.43\% & 39.93\% & 53.39\% \\
    \hline
    70      & 36.50\% & 40.70\% & 55.93\% \\
    \hline
    80      & 39.52\% & 41.52\% & 58.05\% \\
    \hline
    90      & 44.02\% & 42.73\% & 60.06\% \\
    \hline
    100     & 51.14\% & 43.80\% & 61.66\% \\
    \hline
    \end{tabular}
  \label{tab3}
\end{table}

\renewcommand\arraystretch{1.2}
\begin{table}[!htbp]
  \centering
  \caption{The test accuracy of the encrypted model after model pruning attack.}
    \begin{tabular}{|c|c|c|c|}
    \hline
    \multicolumn{1}{|c|}{Pruning rate} & Fashion-MNIST & CIFAR-10 & GTSRB \\
    \hline
    0.1      & 10.38\% & 10.93\% & 6.87\% \\
    \hline
    0.2      & 10.96\% & 10.84\% & 6.95\% \\
    \hline
    0.3      & 11.57\% & 10.60\% & 7.81\% \\
    \hline
    0.4      & 17.65\% & 11.73\% & 7.23\% \\
    \hline
    0.5      & 9.97\% & 15.51\% & 5.94\% \\
    \hline
    0.6      & 10.31\% & 13.08\% & 5.86\% \\
    \hline
    0.7      & 10.62\% & 10.70\% & 5.95\% \\
    \hline
    0.8      & 10.00\% & 12.70\% & 8.21\% \\
    \hline
    0.9      & 10.00\% & 14.80\% & 6.29\% \\
    \hline
    \end{tabular}
  \label{tab4}
\end{table}

\subsubsection{\textbf{Robustness against Model Pruning Attack}}
Model pruning \cite{han2015learning} is an efficient technique to compress DNN model by deleting the redundant parameters with small absolute values.
By implementing the model pruning \cite{han2015learning} technique, an attacker may delete the encrypted parameters.
In the experiment, the pruning rate is set to be 10\%$\sim$90\%, and the experimental results are presented in Table \ref{tab4}.
It can be seen that, for all the three datasets, the test accuracies of the encrypted model are still lower than 20\% after model pruning attack, which demonstrate the robustness of the proposed method against model pruning attack.
The reason is that, the weights of the encrypted parameters are all in a normal range.
To remove the encrypted parameters, other parameters that are not encrypted will be deleted at the same time.
Thus, even if the encryption parameters are deleted, the test accuracy is still at a low level.

\subsubsection{\textbf{Robustness against Adaptive Attack}}
In this section, the worst-case scenario is considered, where the malicious attacker knows the detailed steps of the proposed method, and attempts to conduct the powerful adaptive attack to remove the added adversarial perturbations without the secret key.
In practice, the internal parameters of DNNs are not publicly accessible, which makes the attacker don't have access to the encrypted parameters of the DNN model.
However, in this experiment, we assume that the attacker have the knowledge of all the internal parameters of the DNN model.
Only the key (i.e., the positions of the encrypted parameters or the corresponding added perturbations) is unknown to the attacker.

The attacker can search for the potential encrypted parameters by following the steps described in Section \ref{Encryption}, and calculates the value of the perturbation with Equation \eqref{eq4}, then removes the added perturbations of these parameters to decrypt the model.
The test accuracy of the encrypted model after the adaptive attack is shown in Table \ref{tab5}.
The test accuracy of the model still remains in a low level (56.28\%, 62.03\%, and 63.92\% for Fashion-MNIST, CIFAR-10, and GTSRB, respectively) after the adaptive attack, which demonstrates the robustness of the proposed method against adaptive attack.
Overall, under the adaptive attack where the attackers have knowledge of the mechanism of the proposed method and the internal parameters of the encrypted model, the robustness of the proposed method is still demonstrated.

\renewcommand\arraystretch{1.2}
\begin{table}[!htbp]
  \centering
  \caption{The test accuracy of the encrypted model after adaptive attack.}
    \begin{tabular}{|c|c|}
    \hline
    \multicolumn{1}{|c|}{Dataset} & Test accuracy \\
    \hline
    Fashion-MNIST (DenseNet)  & 56.28\% \\
    \hline
    CIFAR-10 (ResNet)      & 62.03\% \\
    \hline
    GTSRB (AlexNet)         & 63.92\% \\
    \hline
    \end{tabular}
  \label{tab5}
\end{table}

\subsection{\textbf{Comparison with Related Works}} \label{Comparison}
In this section, the proposed method is compared with the existing active DNN IP protection works \cite{pyone2020training}, \cite{ChenW18}, \cite{FanNC19}, \cite{chakraborty2020hardware}.
These related works are introduced as follows.
Pyone \textit{et al.} \cite{pyone2020training} utilize a secret key to preprocess the training images, and the model will be trained on these preprocessed training images.
Unless the input images are preprocessed with the secret key, the model will not output the correct predictions.
Chen and Wu \cite{ChenW18} train an anti-piracy model from scratch, where the model outputs the correct prediction only when the input data is processed by a transformation module.
Fan \textit{et al.} \cite{FanNC19} embeds passports into the middle layers of DNN model.
Without these passports, the model will be dysfunction.
Chakraborty \textit{et al.} \cite{chakraborty2020hardware} utilize the key-dependent algorithm \cite{chakraborty2020hardware} to train a DNN model, which can only work normally in the specific hardware device.

The comparison results are presented in Table \ref{tab6}.
As shown in Table \ref{tab5}, for authorized users, the test accuracy of the five works are all in a high level (around 90\%).
For the three datasets, the accuracy drop $A_d$ of the proposed method is as high as 87.91\%, which is higher than that of works \cite{pyone2020training}, \cite{FanNC19}, \cite{chakraborty2020hardware}, and similar to that of work \cite{ChenW18}.
To realize the function of authorization control, the proposed method only needs to perturb a small amount of parameters of a trained model.
However, for works \cite{ChenW18}, \cite{FanNC19}, \cite{chakraborty2020hardware}, the implementations of these three related works are time-consuming as these works require to retain the DNN model, by which the original parameters of the model will be changed completely.
Moreover, the work \cite{chakraborty2020hardware} requires the support of hardware devices, which is costly in the commercial applications.
Thus, compared with works \cite{ChenW18}, \cite{FanNC19}, \cite{chakraborty2020hardware}, the proposed method require low computational overhead as the proposed method does not require to retrain the DNN models, which also makes the modification of model parameters negligible.
In addition, the proposed method does not require the support of hardware, which is low-cost and feasible in the realistic commercial applications.

\begin{table*}[!htbp]
  \centering
  \caption{Comparison between the proposed method and other active DNN IP protection works}
    \begin{tabular}{|c|c|c|c|c|c|c|c|}
    \hline
    \multicolumn{1}{|c|}{\multirow{2}{*}{Works}} & \multirow{2}{*}{Dataset} & \multicolumn{2}{c|}{Accuracy} & \multicolumn{1}{c|}{\multirow{2}{1.1cm}{\centering Accuracy \\drop $A_d$}} & \multicolumn{1}{c|}{\multirow{2}{2.1cm}{\centering Require\\ additional training}} & \multicolumn{1}{c|}{\multirow{2}{2.1cm}{\centering Require\\ hardware support}} \\
\cline{3-4}       &    & Authorized usage & Unauthorized usage &    &    &      \\
    \hline
    \multicolumn{1}{|c|}{\multirow{3}{1.6cm}{\centering The proposed\\ method}} & Fashion-MNIST & 91.01\% & 10.36\% & 80.65\%  & \multirow{3}{*}{\centering No} & \multirow{3}{*}{\centering No} \\
\cline{2-5}       & CIFAR-10 & 92.02\% & 10.86\% & 81.16\%  &    &  \\
\cline{2-5}       & GTSRB & 94.85\% & 6.94\% & 87.91\%  &    &  \\
    \hline
    \multirow{1}{*}{\centering \cite{pyone2020training}}  & CIFAR-10 & 92.26\% & 20.01\% & 72.25\% & Yes & No \\
    \hline
    \multirow{2}{*}{\centering \cite{ChenW18}} & Fashion-MNIST & 92.55\% & 1.55\% & 91.00\%  & \multirow{2}[3]{*}{Yes} & \multirow{2}[3]{*}{No} \\
\cline{2-5}       & CIFAR-10 & 90.61\% & 0.78\% & 89.83\% &   &  \\
    \hline
    \multirow{1}{*}{\centering \cite{FanNC19}}  & CIFAR-10 & 90.89\% & around 10\% & 80.89\% & Yes & No \\
    \hline
    \multirow{2}{*}{\centering \cite{chakraborty2020hardware}} & Fashion-MNIST &  89.93\% & 10.05\% & 79.88\% & \multirow{2}[3]{*}{Yes} & \multirow{2}[3]{*}{Yes} \\
\cline{2-5}       & CIFAR-10 &  89.54\% & 9.37\% & 80.17\% &   &  \\
    \hline

    \end{tabular}%
  \label{tab6}%
\end{table*}%

\section{Conclusion} \label{Conclusion}
This paper proposes an active DNN IP protection method based on adversarial perturbations.
The proposed method realizes the function of usage control, where the DNN model only works for authorized users, thus prevents the infringement in advance.
Compared with the existing active DNN IP protection works, the proposed method only needs to encrypted an extremely small number of parameters, and does not require additional training or hardware support, which is low-cost and more practical in the commercial applications.
The effectiveness of the proposed method is experimentally demonstrated.
For unauthorized users, the accuracy drops are as high as 80.65\%, 81.16\%, and 87.91\% for Fashion-MNIST \cite{abs-1708-07747}, CIFAR-10 \cite{krizhevsky2009learning} and GTSRB \cite{stallkamp2011german}, respectively.
Besides, the number of the encrypted parameters is as low as 23, and the weights of the encrypted parameters are all in a normal range, which makes the encrypted parameters undetectable for malicious attackers.
In addition, the proposed method is robust against model fine-tuning attack, and model pruning attack, as the test accuracy of DNN model is still in a low level after these two attacks (as low as 61.66\% and 17.64\% after model fine-tuning attack and model pruning attack, respectively).
Moreover, for the adaptive attack where the attackers have knowledge of the detailed steps of the proposed method, the robustness of the proposed method is also demonstrated.
In our future work, the function of user identity management will be further explored.

\bibliographystyle{IEEEtran}
\bibliography{ref}

\end{document}